\documentclass[preprint,aps,pre,amsmath,amssymb]{revtex4}
\usepackage{graphics}
\usepackage{epsf,graphicx,pstcol,epsfig,psfrag}
\begin{document}

\title{From modeling of political opinion formation to two-spin statistical physics model}
\author{Jozef Sznajd}
\affiliation{Institute for Low Temperature and Structure Research, Polish Academy of Sciences, 50-422 Wroclaw, Poland} 
\date{\today}

%%%%%%%%%%%%%%%%%%%%%%%%%%%%%%%%%%%%%%%
\begin{abstract}
A two-spin system based on the idea used to formulate the dynamical model simulating the process of opinion spreading in a society, where people adopt one of the four social attitudes [Physica A {\bf 351}, 593 (2005)] is studied. The model is made of chains with two species of spins $S$ and $V$ at each site coupled by an insite interaction $\tilde M_0$  and interacting in different way with nearest (NN) and next-nearest (NNN) neighbors. The one-dimensional model with only NN interaction of the $S$ spins and both NN and NNN interactions of the $V$ spins (ACLS model) is analyzed. Using the linear renormalization group (LRG) the dependences of the specific heat and correlation functions on temperature and $\tilde M_0$ are calculated.  The roles of the insite and NNN interactions are considered. The usefulness of the ACLS model to define stable states of the dynamical model describing opinion spreading in a society, which members represent four different attitudes to the level of freedom in two areas: personal and economic is discussed. It is pointed out that the difficulties in organizing modern societies can result from an internal conflict between these two areas.
\end{abstract}
\maketitle
%%%%%%%%%%%%%%%%%%%%%%%%%%%%%%%%%%%%%%%%%%%%%%%%%%%

\section{Introduction}

In 2005 we proposed a model to simulate the process of opinion formation in a society, whose members represent four different attitudes instead of traditional two: "leftist" and "rightist" \cite{SW1}. The main assumption was that the attitude space can be divided into two areas, which we called personal and economic and the mechanisms of
opinion formation in both of these areas are different. So, each member of the society (agent) can be characterized by two traits describing the attitude to personal freedom and to economic freedom, respectively.  Each agent can have one of two possible opinions $(+)$ or $(-)$ at each area and tries to influence its neighbors. Additionally, we assumed that in the personal area the information flows inward from the neighborhood (like in most opinion dynamic models \cite{Holyst}), whereas in the economic area the information flows outward to the neighborhood (like in the model proposed by us in the paper Ref.~\onlinecite{SW2} ).  Finally, we have four-state model where each state represents one of the four possible groups of the society: Authoritarian $(- -)$, Conservative $(- +)$, Libertarian $(+ +)$, and Socialist $(+ -)$ (ACLS model). 
The mutual influence of both aspects (personal and economic) was included by a factor of tolerance $p$, which described the probability how the disagreement of the two agents in one area influence their convincing force in the other area.  In Ref.~\onlinecite{SW1}, as often in sociophysics models, the environment was represented by one- and two-dimensional lattices.This sociophysics model bear in on a statistical physics model with two Ising spins at one site. The idea of such a model was proposed by Ashkin and Teller (AT) \cite{AT} as a generalization of the Ising model to describe four-component system. However, the form of the AT Hamiltonian is essentially different from that appropriate for the sociophysics system described above and in one dimension our statistical physics two-spin (TS) model deriving from the sociophysics considerations can be written as follows

\begin{equation}
H_{ACLS} =-\tilde J_1\sum_i  S_i S_{i+1} -\tilde K_1\sum_{i}  V_i V_{i+1} - \tilde K_2\sum_{i}  V_i V_{i+2} -\tilde M_0 \sum_i S_i V_{i}. 
\end{equation}
where $S_i = \pm 1$ and $V_i = \pm 1$ denote two species of Ising spins,  $\tilde J_1$ refers to convincing force in the personal area (standard Ising interaction between nearest neighbors), $\tilde K_i$  refers to convincing forces in the economic area, and $\tilde M_0$ describes a mutual influence of the both areas.

In the present work we study properties of the sociologically motivated model described by the Hamiltonian (1) which can be used to describe thermodynamics of e.g.  dimer or double-spin-Ising chain physical systems \cite{Wen} as well as stationary states of  "complex" systems. 
\section{ACLS model}

ACLS model defined by the Hamiltonian (1)  with the reduced coupling parameters 
\begin{equation}
\tilde j_1 =\frac{\tilde J_1}{k_B T} = \frac{j_1}{T}, \quad \tilde k_1=\frac{\tilde K_1}{k_B T} = \frac{k_1}{T}, \quad  \tilde k_2=\frac{\tilde K_2}{k_B T} = \frac{k_2}{T}, \quad  \tilde m_0=\frac{\tilde M_0}{k_B T}=\frac{m_0}{T}.
\end{equation}
reads
\begin{equation}
H_{ACLS} =\tilde j_1\sum_{i}  S_i S_{i+1} +\tilde k_1\sum_{i}  V_i V_{i+1}+\tilde k_2\sum_{i}  V_i V_{i+2} +\tilde m_0 \sum_i S_i V_{i},
\end{equation} 
as usual \emph {a factor $-\beta = -1/k_B T$ has been absorbed in the Hamiltonian (3)}.

The linear renormalization group (LRG) transformation (decimation) for the Hamiltonian (3) is defined  by
\begin{equation}
e^{H'_R}=Tr_{S,V} P(\sigma, \upsilon; S,V) e^{H_{ACLS}}.
\end{equation}
The weight operator $P(\sigma, \upsilon; S,V) $ which couples the original spins $S, V$ and effective $\sigma, \upsilon$ is chosen in the linear form
\begin{equation}
P(\sigma, \upsilon; S,V)=\prod_{i=0}^{N} p_i(\sigma, \upsilon; S,V)
\end{equation}
and
\begin{equation}
 p_i(\sigma, \upsilon; S,V) =\prod_{n=0}^{\omega} (1+\sigma_{i+n} S_{i+nm})(1+\upsilon_{i+n} V_{i+nm}).
\end{equation}
The next step is the choice of a renormalized block or, in other words, values of $m$ and $\omega$ (6). The smallest nontrivial block that allows to consider both ferromagnetic and antiferromagnetic ground state structure is 4-site one, which means $m=3$. However in order to apply LRG to a model with the second nearest neighbor interaction one should consider at least seven sites (Fig.1), it means $\omega =2$ and the linear RG projector has the form
\begin{equation}
 p_i(\sigma, \upsilon; S,V) = (1+\sigma_1 S_1)(1+\sigma_{2} S_{4})(1+\sigma_{3} S_{7})(1+\upsilon_1 V_1)(1+\upsilon_{2} V_{4}) (1+\upsilon_{3} V_{7}).
\end{equation}

Now in each LRG step three renormalized sites (denoted by squares in Fig.1) survive and the seven site block can realized $4^7$ states and therein $121$ nonequivalent ones. The RG transformation generate all possible interactions acceptable by the symmetry of the problem. In this case 12 new interactions $( \tilde j_2, \tilde k_1, \tilde m_1, \tilde m_2,\tilde  k_{3s}, \tilde k_{3v},\tilde k_{sv}, \tilde k_{sv1}, \tilde k_{sv2}, \tilde k_{vs2}, \tilde k_{sv3}, \tilde k_6)$ come into play, four two-spin, seven four-spin and one six-spin. To be complete one has to add these new interactions to the original Hamiltonian $H_{ACLS}$ (3).
Finally, the renormalization group transformation with the projector (5,6) must be applied to the following Hamiltonian 

\begin{figure}
\label{Fig_1}
 \epsfxsize=8cm \epsfbox{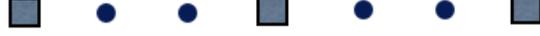}
 \caption{Block used to renormalize the ACLS model (3). Squares denote renormalized and circles decimated spins.}
 \end{figure}

\begin{eqnarray}
H_{r} &=&\tilde j_1\sum_{i}  S_i S_{i+1}+\tilde k_2\sum_{i}  V_i V_{i+2} +\tilde m_0 \sum_i S_i V_{i} \nonumber \\
&+& \tilde j_2\sum_{i}  S_i S_{i+2} +\tilde k_1\sum_{i}  V_i V_{i+1} +\tilde m_1 \sum_i S_i V_{i+1}+\tilde m_2 \sum_i S_i V_{i+2}  \nonumber \\
&+&\tilde k_{3s} \sum_{i}  S_i S_{i+1} S_{i+2}(V_i+V_{i+1} +V_{i+2})+\tilde k_{3v} \sum_{i}  (S_i+S_{i+1} +S_{i+2}) V_i V_{i+1} V_{i+2} \nonumber \\
&+&\tilde k_{sv}  \sum_{i} S_i S_{i+1}  V_i V_{i+1} +\tilde k_{sv1} \sum_{i}  (S_i S_{i+1}  V_{i+1} V_{i+2}+S_{i+1} S_{i+2}  V_{i} V_{i+1}) \nonumber \\
&+&\tilde k_{sv2} \sum_{i}  (S_i S_{i+1}+S_{i+1} S_{i+2}) V_i V_{i+2}+
          \tilde k_{vs2} \sum_{i}  S_i S_{i+2} (V_i V_{i+1}+V_{i+1} V_{i+2})  \nonumber \\
&+&\tilde k_{sv3} \sum_{i} S_i S_{i+2} V_i V_{i+2} 
+\tilde k_6  \sum_{i}  S_i S_{i+1} S_{i+2}  V_i V_{i+1} V_{i+2}.
\end{eqnarray}
For the seven site block (Fig.1) the renormalized Hamiltonian has the form
\begin{eqnarray}
H'_R &=& \log Tr_{S,V} 1/2^{13} (1+\sigma_1 S_1)(1+\upsilon_1 V_1)(1+\sigma_2 S_4)(1+\upsilon_2 V_4)(1+\sigma_3 S_7)(1+\upsilon_3 V_7) e^{H_r}  \nonumber \\
&=& Z_0 +J_1 (\sigma_1 \sigma_2+\sigma_2 \sigma_3)+K_2 (\upsilon_1 \upsilon_{2}+\upsilon_2 \upsilon_3)+ M_{01} (\sigma_1 \upsilon_1+\sigma_3 \upsilon_3 )+M_{02} \sigma_2 \upsilon_2 \nonumber \\
&+& J_2 \sigma_1 \sigma_3 + K_1 \upsilon_1 \upsilon_{3} + M_{1v} (\sigma_1+\sigma_3) \upsilon_2+M_{1s} \sigma_2 (\upsilon_1+\upsilon_3)+M_2 (\sigma_1 \upsilon_3+\sigma_3 \upsilon_1) \nonumber \\
&+& K_{3s} \sigma_1 \sigma_2 \sigma_3 (\upsilon_1+\upsilon_2+\upsilon_3)+K_{3v} (\sigma_1+\sigma_2+\sigma_3) \upsilon_1 \upsilon_2 \upsilon_3 + K_{sv} (\sigma_1 \sigma_2 \upsilon_1 \upsilon_2+\sigma_2 \sigma_3 \upsilon_2 \upsilon_3) \nonumber \\
&+& K_{sv1} (\sigma_1 \sigma_2 \upsilon_2 \upsilon_3+\sigma_2 \sigma_3 \upsilon_1 \upsilon_2)+K_{sv2}  (\sigma_1 \sigma_2 +\sigma_2 \sigma_3)\upsilon_1 \upsilon_3 +K_{vs2} \sigma_1 \sigma_3 (\upsilon_1 \upsilon_2+\upsilon_2 \upsilon_3) \nonumber \\
&+& K_{sv3} \sigma_1 \sigma_3 \upsilon_1 \upsilon_3 + K_6 \sigma_1 \sigma_2 \sigma_3 \upsilon_1 \upsilon_2 \upsilon_3.
\end{eqnarray}

It is not very difficult to find analytical relations between the renormalized parameters ($J_i, K_i, M_i$) (9) and original ones ($\tilde j_i, \tilde k_i, \tilde m_i$) (8), although the final closed expressions contain many terms. It should be noted that the choice of the transformation with the projector (7) drives the inequivalence of the sites $(1, 3)$ and $2$, which leads to two parameters ($M_{01}$, $M_{02}$) in place of original $\tilde m_0$ and ($M_{1v}, M_{1s}$) instead of $\tilde m_1$.
In the following we assume arbitrary $\tilde m_0 \rightarrow M_0 = M_{01}$ and  $\tilde m_1 \rightarrow M_1 = \frac{1}{2}( M_{1v}+M_{1s}) $ and confine ourselves only to bilinear terms. Finally, the renormalized Hamiltonian reads

\begin{eqnarray}
H'_R &=& Z_0 +J_1 (\sigma_1 \sigma_2+\sigma_2 \sigma_3)+K_2 (\upsilon_1 \upsilon_{2}+\upsilon_2 \upsilon_3)+ M_{0} (\sigma_1 \upsilon_1+\sigma_3 \upsilon_3 + \sigma_2 \upsilon_2 )+J_2 \sigma_1 \sigma_3 \nonumber \\
&+& K_1 \upsilon_1 \upsilon_{3} + M_1 [(\sigma_1+\sigma_3) \upsilon_2+ \sigma_2 (\upsilon_1+\upsilon_3)]+M_2 (\sigma_1 \upsilon_3+\sigma_3 \upsilon_1), 
\end{eqnarray}
and the RG transformation has the form of the seven recursion relations
\begin{equation}
(\tilde j_1,\tilde  j_2, \tilde k_1, \tilde k_2, \tilde m_0, \tilde m_1, \tilde m_2) \quad \rightarrow \quad (J_1, J_2, K_1, K_2, M_{01}, (M_{1v}+M_{1s})/2, M_2).
\end{equation}
The explicit form of the recursion relations set is presented in the Appendix Eqs. (25). 

\begin{figure}
\label{Fig_2}
 \epsfxsize=10cm \epsfbox{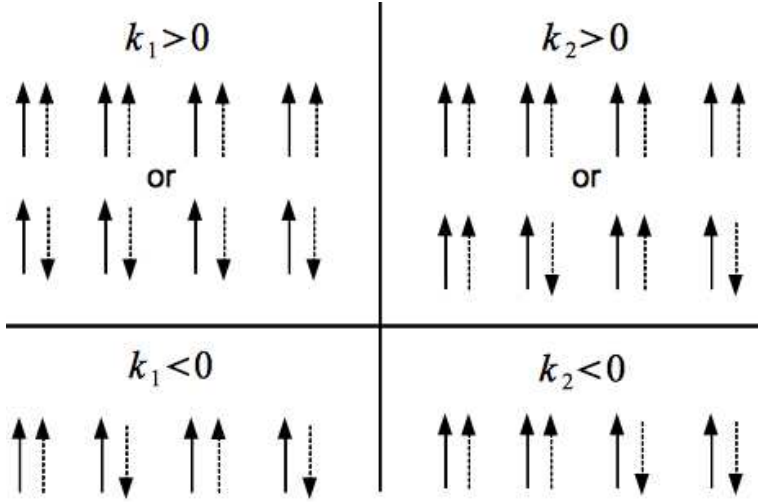}
 \caption{Ground state spin structures of the ACLS model with either $k_ 1\neq 0$ or  $k_2 \neq 0$.}
 \end{figure}

\begin{figure}
\label{Fig_3}
 \epsfxsize=17cm \epsfbox{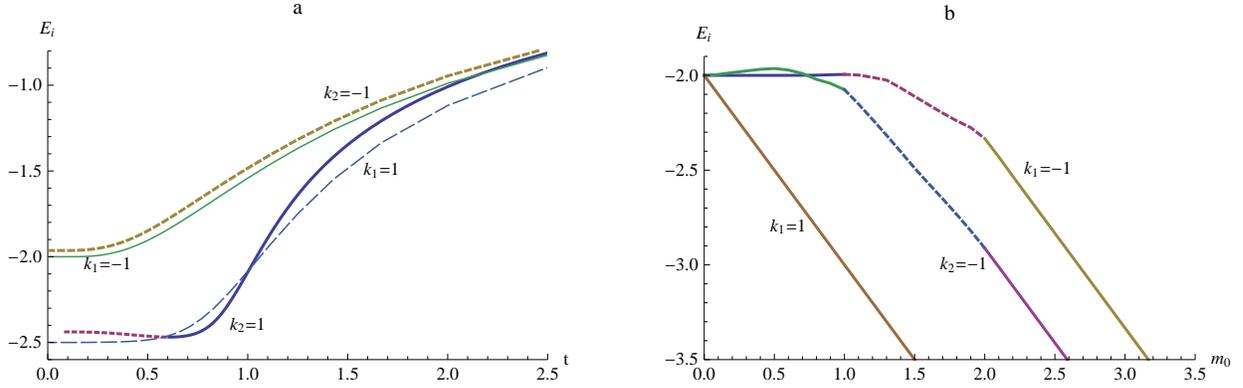}
 \caption{Internal energy as a function of temperature $t$ at $m_0=0.5$ for $k_1=1$ (thin dashed line), $k_1=-1$ (thin solid), $k_2=1$ (thick solid), and $k_2=-1$ (thick dashed) (a) and insite interaction $m_0$ for $k_1=\pm 1$ and $k_2=-1$ at $t=0.1$ (b).}
 \end{figure}

\begin{figure}
\label{Fig_4}
 \epsfxsize=17cm \epsfbox{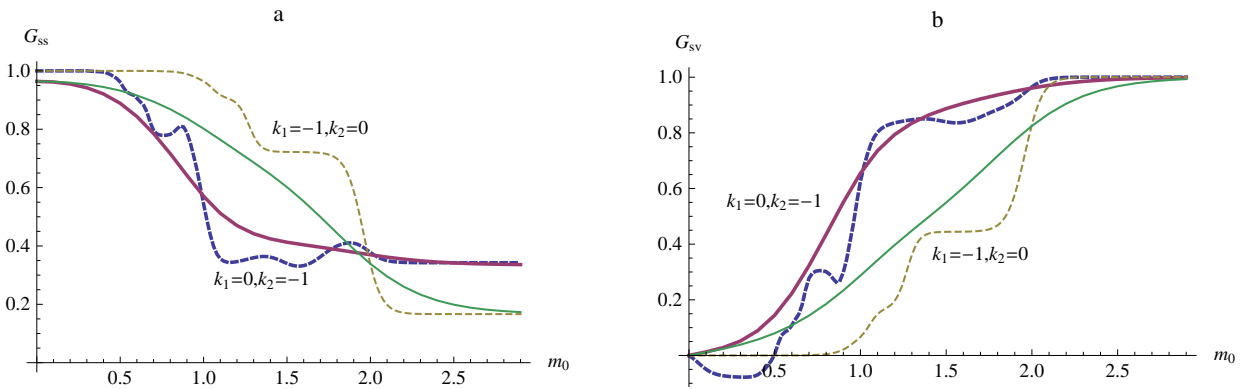}
 \caption{Correlations $G_{ss} = <S_i S_{i+1}>$ (a) and  $G_{sv} = <S_i V_{i}>$ (b) as functions of $m_0$ at  $ t =0.5$ (solid lines) and $t=0.1$ (dashed lines) for two models (i) $k_1=-1, k_2=0$  and (ii) $ k_1=0, k_2  =-1$.}
 \end{figure}

\begin{figure}
\label{Fig_5}
 \epsfxsize=17cm \epsfbox{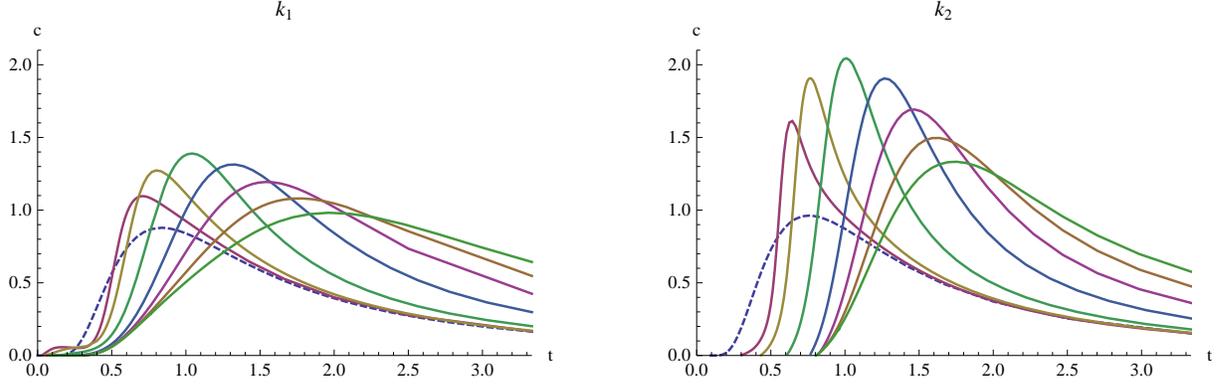}
 \caption{The temperature dependence of the specific heat: ($k_1$) $ k_1 =1$  ($k_2=0$ ) and ($k_2$) $ k_2  =1$  ($k_1=0$) for $m_0=0.1, 0.2, 0.5, 1, 1.5, 2, 2.5$ from left to right and $m_0=0$ (dashed lines).}
 \end{figure}

\begin{figure}
\label{Fig_6}
 \epsfxsize=17cm \epsfbox{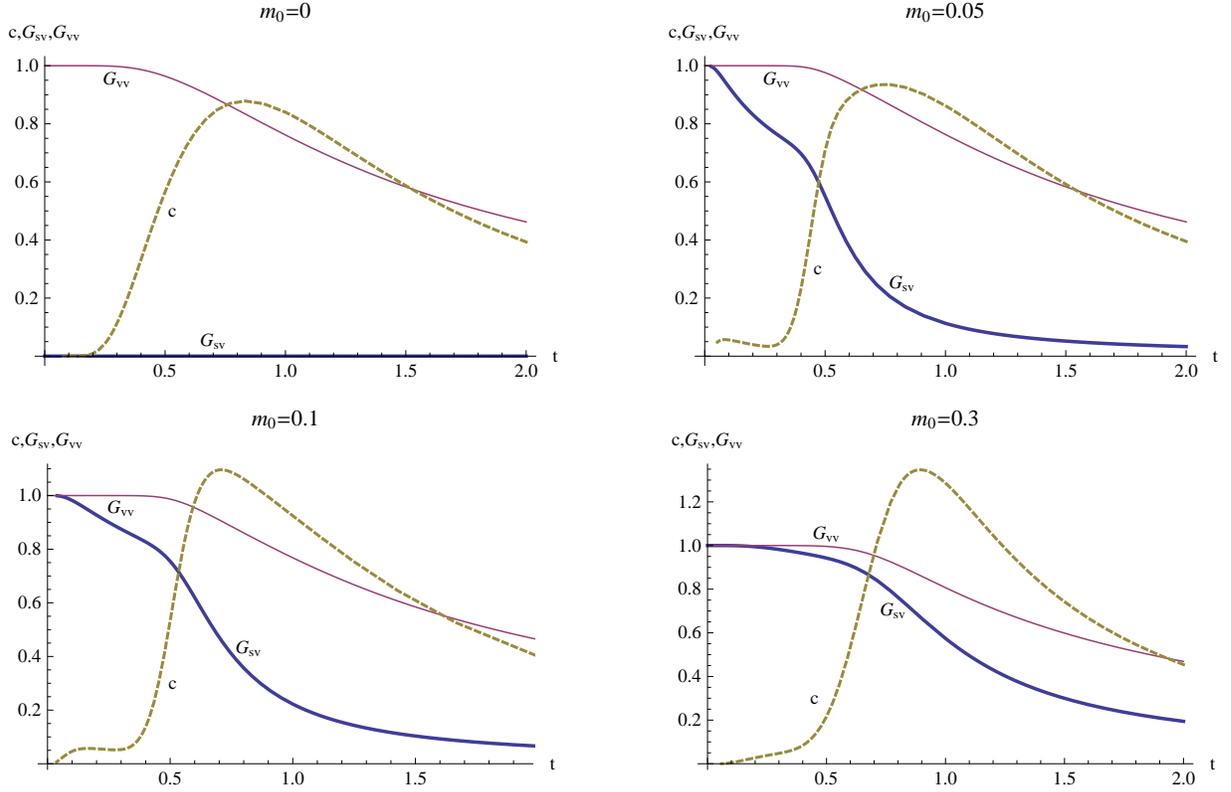}
 \caption{The temperature dependence of the specific heat (dashed lines), intersite correlation $G_{vv}$ (thin), and insite correlation $G_{sv}$ (solid) for $k_1 =1$  ($k_2=0$ ) and $m_0=0, 0.01, 0.05$, and 0.3.}
\end{figure}

\begin{figure}
\label{Fig_7}
 \epsfxsize=17cm \epsfbox{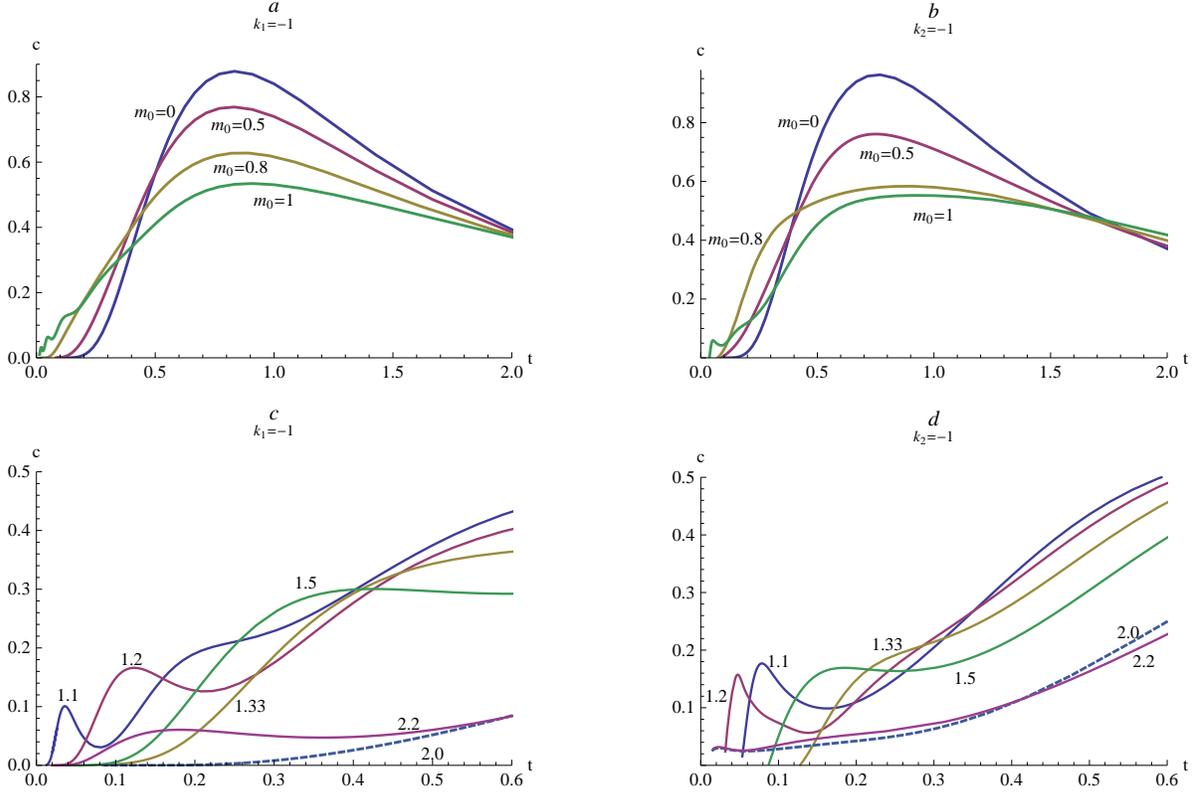}
 \caption{The temperature dependence of the specific heat for $k_1 =-1$ and $k_2=0$  $(a)$ and $(c)$ (left column) and for $ k_2  =-1$ and $k_1=0$ $(b)$ and $(d)$ (right column).}
 \end{figure}

\begin{figure}
\label{Fig_8}
 \epsfxsize=17cm \epsfbox{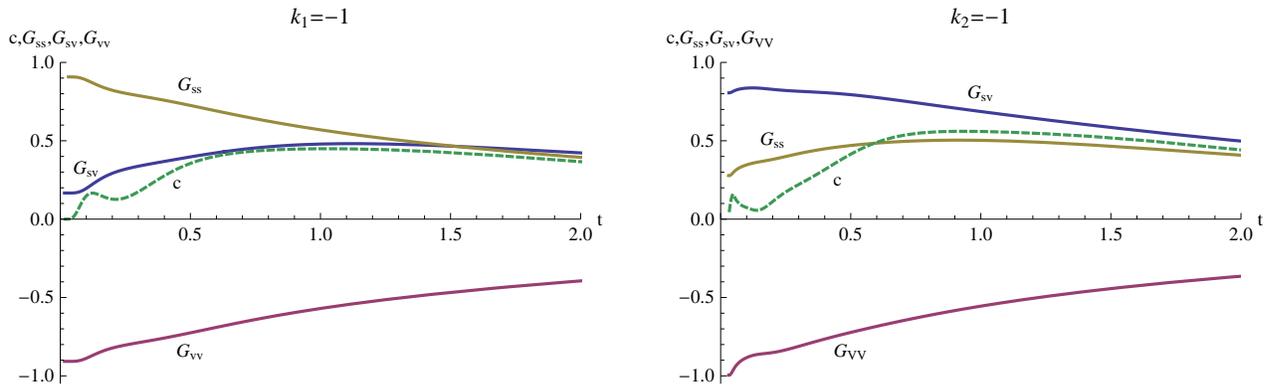}
 \caption{The temperature dependence of the specific heat (dashed lines), intersite correlation $G_{vv}=<V_i V_{i+1}>$ or $G_{VV} = <V_i V_{i+2}>$ and insite correlation $G_{sv}$ (solid lines ) for $k_1 =-1, k_2=0$ (left plot) and $k_2 =-1, k_1=0$ (right plot ) for $m_0=1.2$. }
 \end{figure}

\begin{figure}
\label{Fig_9}
 \epsfxsize=17cm \epsfbox{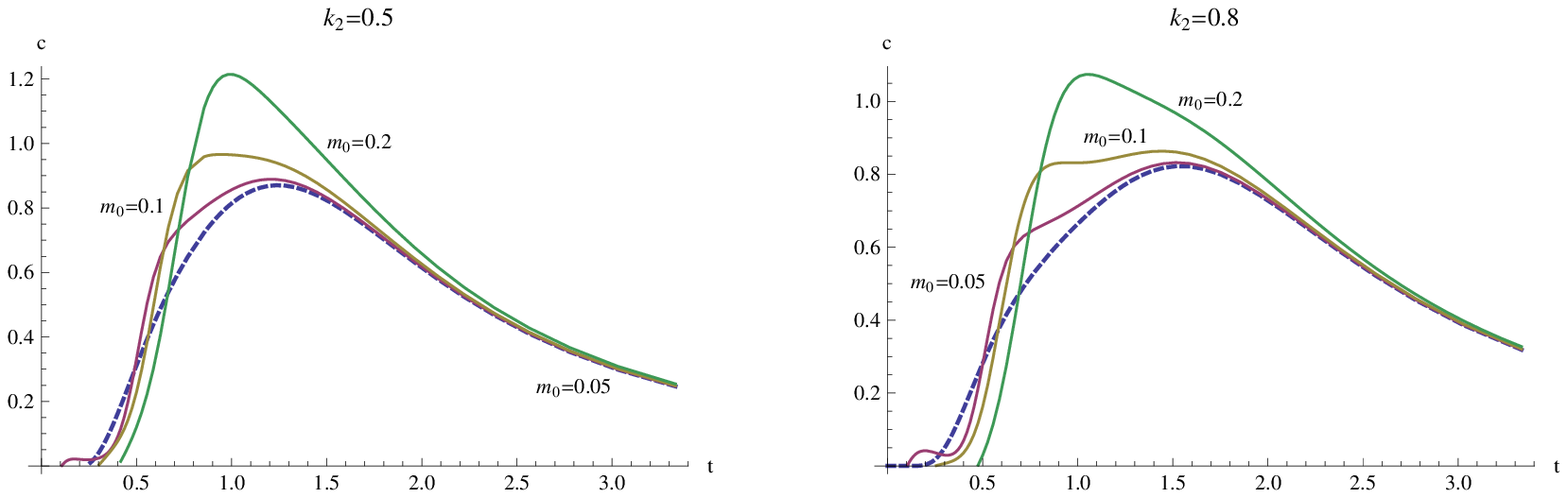}
 \caption{The temperature dependence of the specific heat for $k_1 =1$ and $k_2=0.5$ (left plot) and $ k_2  =0.8$ (right plot) and $m_0=0.05, 0.1,0.2$ and $m_0=0$ (dashed line).}
 \end{figure}

\begin{figure}
\label{Fig_10}
 \epsfxsize=17cm \epsfbox{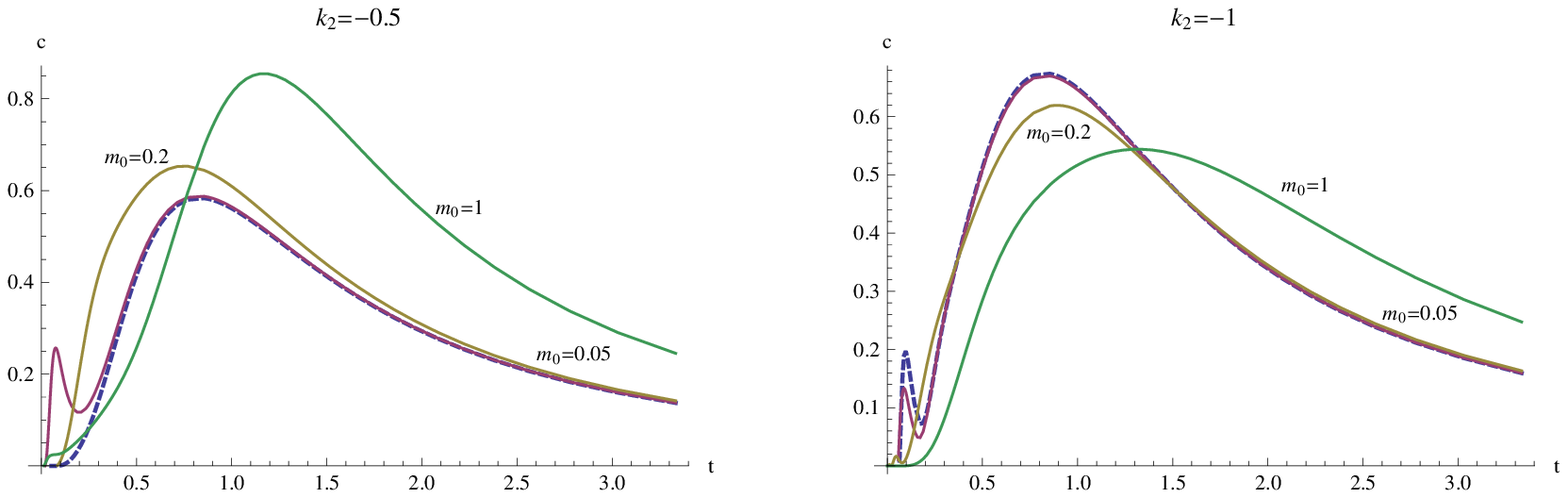}
 \caption{The temperature dependence of the specific heat for $k_1 =1$ and $k_2=-0.5$ (left plot) and $ k_2  =-1$ (right plot) and $m_0=0.05, 0.1,0.2, 1$ and $m_0=0$ (dashed line).}
 \end{figure}

In numerical calculations we will always assume $\tilde J_1 /k_B =j_1= 1$, thus $\tilde j_1 =\tilde J_1/ \beta = t^{-1}$ denotes the reduced inverse temperature. Furthermore we set $j_2=0, m_1=0, m_2=0$ (where $j_i =\tilde j_i T, k_i=\tilde k_i T, m_i=\tilde m_i T$) hence the model parameters are in fact: the $V$ spins interaction  $k_1, k_2$ and the insite coupling $m_0$. Now, we can calculate the free energy per site as a function of temperature collecting the constant terms $Z_0^{(n)}$ generated in each $n$th step of the iteration procedure
\begin{equation}
f =\sum_{n=1}^{\infty} \frac{Z_0^{(n)}}{3^{n}},
\end{equation} 
and then the internal energy $E_i$, specific heat and correlation functions:
\begin{eqnarray}
G_{ss} = <S_i S_{i+1}>, \quad G_{vv} = <V_i V_{i+1}>, \quad G_{VV} = <V_i V_{i+2}>, \quad G_{sv} = <S_i V_i>.
\end{eqnarray}

It is obvious that for a model which exhibits a finite temperature phase transition the RG transformation should preserve the symmetry of the ordered phase structure. But also for the systems that do not undergo such a transition the ground state symmetry impact on finite temperature behavior especially at low temperature. Unfortunately, for some models it is impossible to find a block transformation which preserve all ground state symmetries, e.g. in cases where long-periodic commensurate or incommensurate ground state structures can be expected. This may be an additional reason why LRG, which is still a high temperature method, may fail at low temperatures. In the first version of the original ACLS model (1) only the next-nearest-neighbor interaction is taking into account ($k_2>0, k_1=0$). However, studying the physical systems different cases should be considered. To compare the function of NN and NNN interactions we will analyze two cases: $k_ 1 \neq 0$ ($k_2=0$) and $k_ 2 \neq 0$ ($k_1=0$) but also $k_1=1$ for several values of $k_2$ and finally $k_1 = k_2$.
 
Let us start with the models without insite interaction $m_0=0$. The ground state spin structures for several cases are presented in Fig.2. For $k_1>0$ ($k_2=0$) the system realizes in each subsystem the ferromagnetic state $L$ ($\uparrow \uparrow \quad \uparrow \uparrow$)   {\em or \/} $S$  ($\uparrow \downarrow \quad \uparrow \downarrow$), and for the opposite directed spins $A$ or $C$), which we have named according to the nomenclature adopted in the sociophysics paper \cite{SW1}  (see Introduction),  whereas for $k_2>0$ ($k_1=0$) $L$ ($\uparrow \uparrow \quad \uparrow \uparrow$) {\em or\/} $LS$ ($\uparrow \uparrow \quad \uparrow \downarrow$) (equivalently $A$ or $AC$). For the antiferromagnetic interactions $k_1 < 0$ there is the $LS$ state and for $k_2<0$ doubled $LS$ state $LLSS$ ($\uparrow \uparrow \quad \uparrow \uparrow \quad  \uparrow \downarrow \quad \uparrow \downarrow$).  For a finite insite coupling $m_0 \neq 0$ the zero-temperature  phase diagram becomes more complex and the LRG transformation (4-7) does not allow us to guess the ground state structure.

Figure 3 shows the internal energy $E_i$ as a function of reduced temperature $t$ for $m_0=0.5$ and $k_i=\pm 1$ [Fig.3(a)], and insite coupling $m_0$ for $k_1=\pm 1$ and $k_2=-1$ at relatively low temperature $t=0.1$ [Fig.3(b)]. For $k_1=1$, $E_i =-( j_1+k_1+m_0)$ in the whole range of $m_0$ which suggests, of course, fully ordered ferromagnetic ground state. 
For the antiferromagnetic interaction $k_1 =-1$ the finite temperature behavior suggests several possibilities: for $0 \geq m_0 \geq 1$, $E_i \approx j_1+k_1$ is constant; for  $m_0  \geq 2$, $E_i=-(j_1+k_1+m_1 - \frac{5}{3}$); and for $1 \geq m_0 \geq 2$ the dependence is more complex, whereas for $k_2=-1$ the linear dependence of $E_i$ on $m_0$ is for $m_0 > 1$  [Fig.3(b)]. As seen in Fig.3(a) in contrast to the other cases for $k_2 = 1$ at some
temperature $t = t^{*} (m_0)$, the function $E_i$ ceases to be convex and so for the temperatures
$t < t^{*} (m_0)$ in the present version the LRG method cannot be applied.

In order to have some insight into ground state spin structure one can look into the temperature dependences of the correlation functions. In Fig.4 the intersite $G_{ss} = <S_i S_{i+1}>$ and insite $G_{sv} = <S_i V_i>$ correlations are presented for both considered cases at the temperature $t=0.5$ and $t=0.1$. At the lower temperature clear plateaus are visible related to several ground states. And so, in the case of $k_1=-1$ for $m_0<1$, $G_{ss}=1$ and $G_{sv}=0$ which means the ferromagnetic order in both subsystems $S$ and $V$ and spin singlet at each site. On the other hand for $m_0 > 2$ $G_{sv} = 1$ and the triplet state in each site is expected. For intermediated values of $m_0$ two other plateaus are observed but within the used method the character of the spin structure related to these plateaus cannot be determined. For $k_2 =-1$ the situation is even more complex and except for large values of $m_0 > 2$ for which $G_{sv}=1$ nothing reliable about the ground state spin structure from the finite temperature behavior of the correlation functions can be deduced.

In Fig.5 the temperature dependences of the specific heat for the models with positive $k_i$ are shown. As seen in both cases $k_1=1$ ($k_2=0$) and $k_2=1$ ($k_1=0$) the insite coupling $m_0$ shifts the specific heat maximum first, toward lower temperature and the maximum height increases and then from $m_0 =0.5$ to higher temperature and the hight decreases. However, there are also marked differences among these cases. While, as already mentioned, for $k_2=1$ the LRG fails at low temperature and leads to non-physical results (negative specific heat), for $k_1=1$ the specific heat tends to zero just like for the standard Ising model provided that $0 <m_0 \leq m^*_0 \approx 0.2$. 
In this region an additional small low-temperature maximum is observed accompanied by an additional inflection point of the insite correlation function $G_{sv}$ (Fig.6), while the intersite correlation $G_{vv} =G_{ss}$ is smooth in this region. So, we would guess that the low temperature specific heat hump marks the temperature at which the spins $S$ and $V$ at the same site become unbound. For higher value of $m_0 \ge 0.3$ this low temperature hump disappears.

Fig.7 shows the temperature dependences of the specific heat for the models with antiferromagnetic coupling in the $V$ spins subsystem $k_1=-1$ or $k_2=-1$. Figs.7(a) and 7(b) display the curves for $m_0 \leq 1$ and Figs.7(c) and 7(d) for $1 < m_0 \leq 2.2$. As seen in both cases  for $m_0$ around $1$ the specific heat behaves erratically caused by the proximity of the frustration point. For $m_0 < 1$ and $k_1=-1$ the specific heat curve tends to zero as for the standard Ising model, whereas for $k_2=-1$ the shape of the specific heat curve is a little different and the method fails at very low temperatures. For $1<m_0<m_0^{max} \approx 1.3$ the specific heat exhibits very clear additional maximum at the low temperature which disappears for higher values of $m_0$. As in the previous case of the  ferromagnetic interactions, it is possible to check the origin of this maximum evaluating the correlation functions $G_{sv}$ and $G_{vv}$ or $G_{VV}$ [$G_{VV}$ for the model with $k_2=-1 (k_1 = 0)]$ (13) . As seen in Fig.8 unlike the ferromagnetic case ($k_i =1$) now the maximum of the specific heat is accompanied by the inflection points of all correlation functions.

Admittedly, no evidence of a real one-dimensional physical system  with only next-nearest-neighbor interactions ($k_1=0, k_2 \neq 0$) is available, it seems to be meaningful to speculate about what will happen in such a situation.
However, in  a physics context, more realistic is, of course, a model with both finite couplings \cite{Wen}. We now consider a few such examples. In Figures 9 and 10 the temperature dependences of the specific heat for $j_1=1, k_1 =1$,  $k_2 = 0.5, 0.8$ (Fig.9) and $k_2=-0.5, -1$ (Fig.10) and several values of the insite interaction $m_0$ are shown.
As seen in Fig.9 for positive $k_2$ just like in the case of $k_2=0$ a slight hump for small $m_0$ is visible at low temperature. Additionally, at higher temperature a second "bulge" appears clearly visible for larger value of $k_2$.
For negative $k_2$ (Fig.10) the sharp peak is observed at low temperature for  a sufficiently small $m_0$. 
Such a double peak specific heat structure was observed e.g. in $SrHo_2O_4$ \cite{Wen}.

\section{Conclusion}
In this paper, we have considered two-spin model inspired by the idea we used to describe social dynamics of the four-party political system \cite{SW1}. The model consists of two species of spins $S_i$ and $V_i$ localized in each site $"i"$. The $S_i$ spins are coupled by the ferromagnetic nearest-neighbor interaction ($\tilde J_1$) and $V_i$ spins by ferro- or anti-ferromagnetic nearest-neighbor $\tilde K_1$ or/and next-nearest-neighbor $\tilde K_2$ interactions. Additionally, the two kind of spins are coupled by the insite interaction $M_0$. The first motivation for the study was to test the applicability of the linear renormalization group technique to the description of the thermodynamic properties of the physical compounds such as double-spin-chain systems with NNN interactions \cite{Wen}. The second motivation was to analyze the statics of the model which is a counterpart of the sociophysics model with two kinds of information flows: from the initial pair of agents to their neighbors in one area of the attitudes (economic area) - outward flow, and from neighbors to agents (personal area) - inward flow. It has been claimed \cite{BS} that, in one dimension, the direction of the information flow is actually irrelevant to the dynamics. The other questionable point in "outward flow model" is an antiferromagnetic rule. It is generally accepted that if the agents of the pair (group) share the same opinion, they successfully impose their opinion on neighbors. However, the rule that if the  agents of the pair disagree, then the nearest neighbor of each agent disagrees with him is usually believed unrealistic. Nevertheless, it seems to be considered a manifestation of anti-conformism. So, we have studied both cases $\tilde K_2>0$ and $\tilde K_2<0$.

{\bf Physics:} The main object of our interest is the role of the insite coupling $m_0$ and the difference between the nearest- and next-nearest-neighbor interactions ($k_1, k_2$) in  the emergence of various behaviors of the ACLS model. For $m_0=0$ and ferromagnetic $k_i$ interactions the ground state is degenerated in both cases $k_1>0$ $(k_2=0)$ and $k_2>0$ $(k_1=0)$, additionally, in the latter case the unit cell is doubled in one of the possible states (Fig.2). The inclusion of insite coupling $m_0$ makes the used block transformation RG fails below relatively high temperature $t = t^{*} (m_0)$ (order of magnitude of $j_1$ interactions for $m_0=0.5$) in this case. For other cases, $k_1>0$ and antiferromagnetic $k_i<0$ interactions the LRG transformation can be used to much lower temperatures. For $k_1>0$ the insite coupling $m_0$ removes the degeneracy but for sufficiently small $m_0$ and at low temperature there is a little hump observed (Fig.5). The temperature range for which this hump occurs is also indicated by an inflection point of the correlation function $G_{sv}$ curve. The other correlations $G_{ii}$ are smooth temperature functions (Fig.6). For $k_1<0$ the LS phase (alternating triplet, singlet)  and for $k_2<0$ the LLSS phase (triplet, triplet, singlet, singlet) at the ground state are realized, and for finite $m_0$ frustration effects come into play. In considered case $k_i=-1$ the system is frustrated for $\mid m_0 \mid =1$ which is seen in the specific heat behavior (Fig.7). For $\mid m_0 \mid > 1$,  in both cases, the additional specific heat maximum appears.  Unlike the ferromagnetic interactions $k_i>0$, in the antiferromagnetic case the low temperature specific heat maximum is accompanied by the inflection points of all correlation functions. This suggests that while for positive $k_i$ the low temperature specific heat hump (Fig.6) marks the temperature at which only the spins at the same site $S_i$ and $V_i$ start to be unbound, the low temperature specific heat maximum for negative $k_i$ is connected with simultaneous unbinding of the insite and neighboring spins. For positive $k_1$ and negative $k_2$ the double maximum specific heat structure is found with sharp low temperature peak observed in some real magnetic compounds \cite{Wen}.

Summarizing, the LRG can be useful to study the finite temperature properties of the ACLS model, however, it fails 
at low temperatures, especially, if the RG transformation does not preserve the ground state symmetries e.g. for the models which exhibit long-periodic or incommensurate ground state structures.

{\bf Sociophysics:} 
For several decades the statistical physics models are used to describe social dynamics, in particular an opinion formation or information spreading \cite{Fort}. Many of these models show the presence of phase transitions. However, it is known that in the social system such a transition does not exist in a thermodynamic sense. Toral and Tessone \cite{TT} have considered the finite size effects in some dynamics opinion formation models and stated that in these models a change of the behavior occurs at some {\emph {pseudo-critical}}  value of the parameter triggering the change (without singularities inherent to the thermodynamic phase transitions) just like it happens in one dimension. So, one can expect some basic qualitative description of the opinion change in the frame of 1D models.

The ACLS model  \cite{SW1} is based on the idea that the area of human interactions can be divided into two subareas: personal and economic. The opinion formation processes is different in these two subareas, and so in the personal area the information flows inward from the neighborhood and in the economic area flows outward to the neighborhood. In the latter case two different dynamic rules were proposed \cite{SW2}. The first rule (r1) is the same in both cases "when members of a pair have the same opinion, then their nearest neighbors agree with them", and we proposed two variants of the second rule (r2): (i) "when members of a pair have different opinions, then the nearest neighbor of each member disagrees with him (her)", (ii) "if the members of a given community are less prone to oppose nearest neighbors, then one should keep first rule (r1) and skip rule (r2)".  In subsequent studies the first variant (i) was considered unrealistic and replaced \cite{Stauffer1, Stauffer2} by various recipes among which the second variant (ii) turned out to be the most popular \cite{Fort}. In  Ref.~\onlinecite{Kasia1} was proposed to control the outflow dynamics (ii) of the one-dimensional Ising-like systems by  using a local "disagreement" function.
\begin{figure}
\label{Fig_11}
 \epsfxsize=17cm \epsfbox{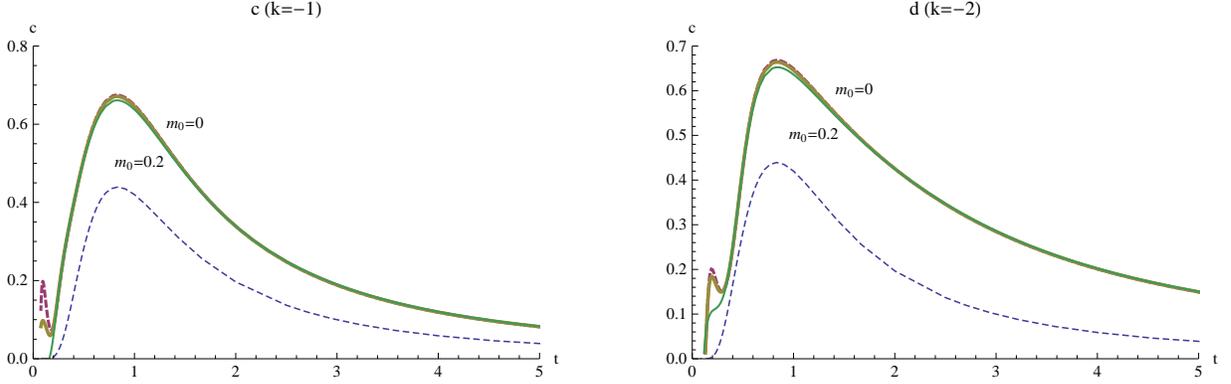}
 \caption{The temperature dependence of the specific heat for $k_1 = k_2 = k =1$ and $2$ and several values of the intersite interaction $m_0$. a) $k=1$, $m_0 =0, 0.05, 0.1, 0.15, 0.2$ and $0.25$ from bottom to the top; b) $k=2$, $m_0 = 0, 0.1, 0.2, 0.3, 0.5$ and $1$ from bottom to the top; c) $k=-1$, $m_0$ = 0 (thick dashed line), 0.1 (thick line) and $0.2$ (thin);  d) $k=-2$, $m_0$ = 0 (thick dashed line), 0.1 (thick line) and $0.2$ (thin);  
Thin dashed lines denote the specific heat of the standard Ising model with $k=0$.}
 \end{figure}
It is easy to see  that such a function can have the following form:
\begin{equation}
E_i=-\tilde K   (V_i +V_{i+1} )(V_{i-1} +V_{i+2} ),
\end{equation}

\begin{figure}
\label{Fig_12}
 \epsfxsize=17cm \epsfbox{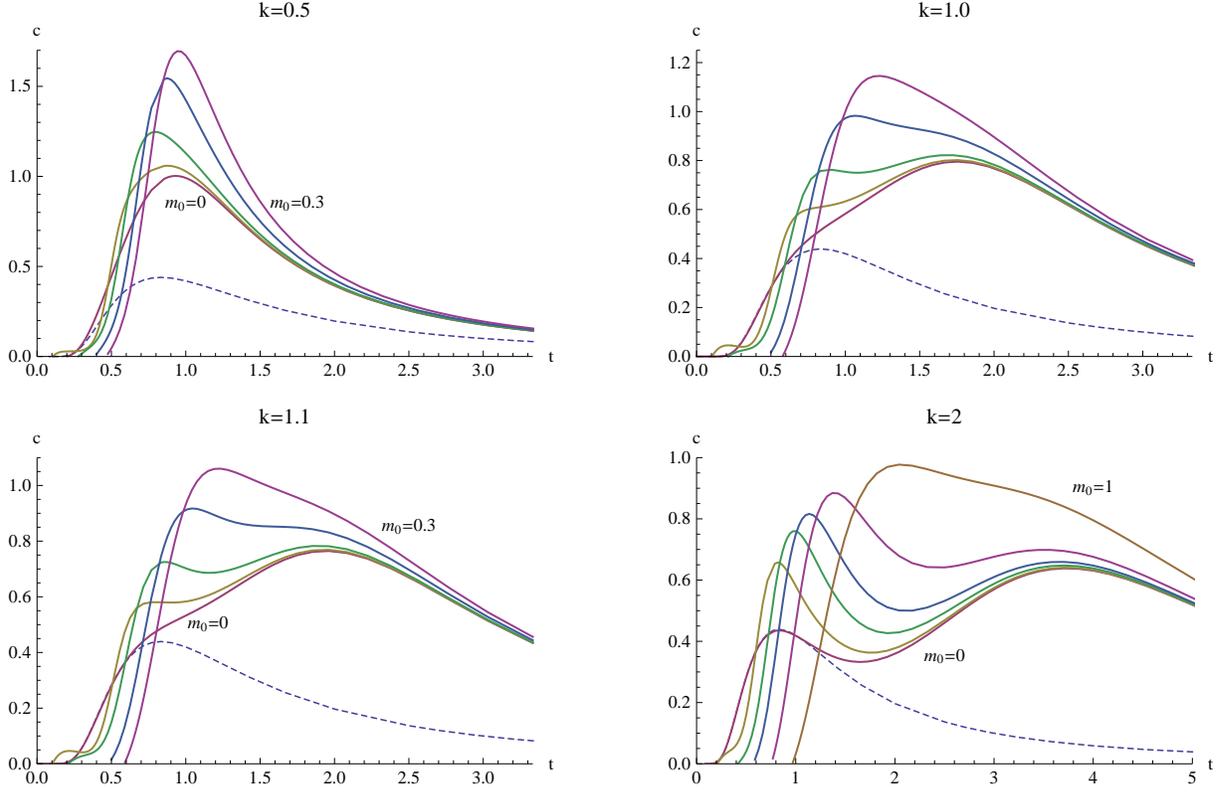}
 \caption{The temperature dependence of the specific heat for $k_1 = k_2 = k =0.5, 0.9, 1.0, 1.1$ and several values of the intersite interaction $m_0 =0,  0.05, 0.1$, and 0.2 from bottom to the top.
Thin dashed lines denote the specific heat of the standard Ising model with $k=0$.}
 \end{figure}

The consensus formation model corresponds to $\tilde K $ positive. However, it seems sensible to consider also $\tilde K$ negative, especially in the processes of forming opinions on economic matters. In the latter case the nearest neighbors disagree with the common opinion of the chosen pair. 
As it was stressed \cite{Fort} the sum of the function $E_i$ over all spins does not play the role of the energy: the local minimization of $E_i$ can lead to an increase of its global value \cite{Kasia2} .
However, if one extend the local function (14) to the whole chain \cite{Galam} then the ACLS Hamiltonian (1) with $\tilde K_1 = \tilde K_2 = \tilde K$ should define the stable states to which converge the four party dynamical model \cite{SW1} for long time. In physics, the crucial quantity which define the state is the temperature. 
Its sociophysical counterpart: "social temperature" quantified, generally speaking, randomness or information noise. This quantity is not universal  and has different meanings in various social processes. For example in politics it can be measure of dissatisfaction with the existing political system. In physics, a lot can be learned about a system by studying a response to various changes of the external parameters. 
In the case of the social temperature an appropriate response function is a counterpart of the specific heat. In politics it should describe the sensitivity of the political system on dissatisfaction of the society members. In Fig.11 the temperature dependences of the specific heat for both cases $k =\frac {\tilde K}{k_B}$ positive and negative and several values of the insite coupling $m_0$ are presented. As seen the influence of $m_0$ on the shape of the specific curve is very different in several cases and for negative $k$ the specific heat is almost $m_0$ independent, except for very low temperature. On the contrary, for positive $k$ which refers to the consensus formation model, both the hight and for larger $k$ also location of the specific heat maximum are very sensitive to the insite coupling $m_0$.  In Fig.12 the temperature dependences of the specific heat for four values of $k = 0.5, 0.9, 1.0$, and $1.1$ and for four values of $m_0 = 0, 0.05, 0.1$, and $0.2$ are presented. As seen in all cases for $m_0=0$ the specific heat goes smoothly to zero, whereas for small $m_0=0.05$ at low temperature the little hump is observed. For larger $m_0$ and $k \ge 0.9$ an additional maximum appears and in all cases the LRG method fails at low temperature, leading to non-physical results. In sociophysics it can signify that e.g. the existing four-party political system is not acceptable any more. It means the getting a long time consensus is difficult or even impossible. 
The point is that in both cases $k$ negative or positive, the insite coupling $m_0$ can trigger or remove a specific heat maximum which can indicate a kind of local (short range) reordering.

It is obvious that the stable ground state in which all members of the society are satisfied is unattainable utopia. Thus, the social temperature plays a crucial role in the description of the society state, and a specific heat maximum denotes growing dissatisfaction which can lead to the change of the political system or in the extreme case to anarchy. Reliable interpretation requires further research, in particular system dynamics. However, when studying the state of  modern societies one should take into account possible internal conflicts between different areas of the attitudes, personal and economic found in the democratic systems, which can lead to instability of such a system.

\section{Appendix}

In this appendix we show derivation of the recursion relations for the coupling parameters of the renormalized Hamiltonian $H'_R$ (10) as functions of the original  parameters 
We start with the Hamiltonian (3)
\begin{eqnarray}
H_{ACLS}(S,V) &=&\sum_{i=1}^5 \tilde j_1 \large[  S_i S_{i+1}+\tilde k_1  V_i V_{i+1}+\tilde k_2 V_i V_{i+2} +\tilde m_0  S_i V_{i} \large]
\end{eqnarray}
Let us denote the RG transformation by ${\cal R}$, then
\begin{equation}
{\cal R} H_{ACLS}(S,V)  \rightarrow H_{r}(\sigma, \upsilon)
\end{equation}
where $H_{r}(\sigma, \upsilon)$ contains all couplings allowed by the symmetry of the problem. To be complete we must add all these generated by the transformation terms to the original Hamiltonian i.e.
\begin{equation}
 H_{ACLS}(S,V)  \rightarrow H_r(S,V)
\end{equation}
and
\begin{eqnarray}
H_{r} (S,V) &=&\sum_{i} \large[\tilde  j_1 S_i S_{i+1}+\tilde k_2 V_i V_{i+2} +\tilde m_0  S_i V_{i} + \tilde j_2  S_i S_{i+2} +\tilde k_1  V_i V_{i+1} 
+\tilde m_1  S_i V_{i+1} +\tilde m_2  S_i V_{i+2}  \nonumber \\
&+&\tilde k_{3s}   S_i S_{i+1} S_{i+2}(V_i+V_{i+1} +V_{i+2})+\tilde k_{3v}  (S_i+S_{i+1} +S_{i+2}) V_i V_{i+1} V_{i+2} \nonumber \\
&+& \tilde k_{sv}   S_i S_{i+1}  V_i V_{i+1} +\tilde k_{sv1}   (S_i S_{i+1}  V_{i+1} V_{i+2}+S_{i+1} S_{i+2}  V_{i} V_{i+1}) \nonumber \\
&+&\tilde  k_{sv2}   (S_i S_{i+1}+S_{i+1} S_{i+2}) V_i V_{i+2}+
           \tilde  k_{vs2}   S_i S_{i+2} (V_i V_{i+1}+V_{i+1} V_{i+2})  \nonumber \\
&+& \tilde k_{sv3}  S_i S_{i+2} V_i V_{i+2} 
+\tilde k_6    S_i S_{i+1} S_{i+2}  V_i V_{i+1} V_{i+2} \large].
\end{eqnarray}
Now we rewrite the RG transformation in the form
\begin{equation}
H_x(\sigma, \upsilon) = Tr_{S,V} p(\sigma, \upsilon; S,V) e^{H_r (S,V)}, \qquad  H'_R = \ln H_x(\sigma,\upsilon) 
\end{equation}
with the seven spin block weight operator
\begin{equation}
 p(\sigma, \upsilon; S,V) =(1+\sigma_1 S_1)(1+\upsilon_1 V_1)(1+\sigma_2 S_4)(1+\upsilon_2 V_4)(1+\sigma_3 S_7)(1+\upsilon_3 V_7) /2^{13}
\end{equation}
the Hamiltonian $H_x$ takes the form
\begin{eqnarray}
H_x &=& H_0 + H_{\sigma12} (\sigma_1 \sigma_2 + \sigma_2 \sigma_3) + H_{\sigma13} \sigma_1 \sigma_3 + H_{\upsilon12} (\upsilon_1 \upsilon_2 + \upsilon_2 \upsilon_3) + 
   H_{\upsilon13} \upsilon_1 \upsilon_3  \nonumber \\
&+& H_{\sigma\upsilon1} (\sigma_1 \upsilon_1 + \sigma_3 \upsilon_3) + H_{\sigma\upsilon2} \sigma_2 \upsilon_2 + 
   H_{\sigma1\upsilon2} (\sigma_1 \upsilon_2 + \sigma_3 \upsilon_2) + H_{\sigma1\upsilon3} (\sigma_1 \upsilon_3 + \sigma_3 \upsilon_1) \nonumber \\
&+& 
   H_{\sigma2\upsilon1} (\sigma_2 \upsilon_1 + \sigma_2 \upsilon_3) + H_{\sigma\sigma\sigma\upsilon1} \sigma_1 \sigma_2 \sigma_3 (\upsilon_1 + \upsilon_3) + 
   H_{\sigma\sigma\sigma\upsilon2} \sigma_1 \sigma_2 \sigma_3 \upsilon_2  \nonumber \\
&+& 
   H_{\sigma1\upsilon\upsilon\upsilon} (\sigma_1 + \sigma_3) \upsilon_1 \upsilon_2 \upsilon_3+ H_{\sigma2\upsilon\upsilon\upsilon} \sigma_2 \upsilon_1 \upsilon_2 \upsilon_3 + H_{\sigma1\sigma2\upsilon1\upsilon2} (\sigma_1 \sigma_2 \upsilon_1 \upsilon_2 + \sigma_2 \sigma_3 \upsilon_2 \upsilon_3) \nonumber \\
&+& 
   H_{\sigma1\sigma2\upsilon2\upsilon3} (\sigma_1 \sigma_2 \upsilon_2 \upsilon_3 + \sigma_2 \sigma_3 \upsilon_1 \upsilon_2) + 
   H_{\sigma1\sigma2\upsilon1\upsilon3} (\sigma_1 \sigma_2 + \sigma_2 \sigma_3) \upsilon_1 \upsilon_3 \nonumber \\
&+& H_{\sigma1\sigma3\upsilon1\upsilon2} \sigma_1 \sigma_3 (\upsilon_1 \upsilon_2 + \upsilon_2 \upsilon_3) +
    H_{\sigma1\sigma3\upsilon1\upsilon3} \sigma_1 \sigma_3 \upsilon_1 \upsilon_3 + H_{\sigma\upsilon6} \sigma_1 \sigma_2 \sigma_3 \upsilon_1 \upsilon_2 \upsilon_3.
\end{eqnarray}
where
\begin{eqnarray}
H_0 &=& H_x \quad for \quad  \sigma_1 \rightarrow 0, \sigma_2 \rightarrow  0, \sigma_3\rightarrow 0, \upsilon_1 \rightarrow 0, \upsilon_2 \rightarrow 0, \upsilon_3\rightarrow  0, \nonumber \\
H_{\sigma12} &=&\frac{H_x-H_0}{\sigma_1 \sigma_2} \quad for \quad \sigma_3\rightarrow 0, \upsilon_1 \rightarrow 0, \upsilon_2 \rightarrow 0, \upsilon_3\rightarrow  0, \nonumber \\
H_{\sigma13} &=&\frac{H_x-H_0}{\sigma_1 \sigma_3} \quad for \quad \sigma_2 \rightarrow  0,  \upsilon_3\rightarrow  0,  \upsilon_3\rightarrow  0,  \upsilon_2 \rightarrow 0, \nonumber \\
H_{\sigma\upsilon1} &=&\frac{H_x-H_0}{\sigma_1 \upsilon_1} \quad for \quad \sigma_2 \rightarrow  0, \sigma_3\rightarrow 0,  \upsilon_2 \rightarrow 0, \upsilon_3\rightarrow  0, \nonumber \\
H_{\sigma\upsilon1} &=&\frac{H_x-H_0}{\sigma_2 \upsilon_2} \quad for \quad  \sigma_1 \rightarrow 0,  \sigma_3\rightarrow 0, \upsilon_1 \rightarrow 0,  \upsilon_3\rightarrow  0, \nonumber \\
H_{\sigma1\upsilon2} &=&\frac{H_x-H_0}{\sigma_1 \upsilon_2} \quad for  \sigma_2 \rightarrow  0, \sigma_3\rightarrow 0, \upsilon_1 \rightarrow 0,  \upsilon_3\rightarrow  0, \nonumber \\
H_{\sigma1\upsilon3} &=&\frac{H_x-H_0}{\sigma_1 \upsilon_3} \quad for \quad   \sigma_2 \rightarrow  0, \sigma_3\rightarrow 0, \upsilon_1 \rightarrow 0, \upsilon_2 \rightarrow 0, \nonumber \\
H_{\sigma2\upsilon1} &=&\frac{H_x-H_0}{\sigma_2 \upsilon_1} \quad for \quad  \sigma_1 \rightarrow 0,  \sigma_3\rightarrow 0,  \upsilon_2 \rightarrow 0, \upsilon_3\rightarrow  0,
\end{eqnarray}
and
\begin{eqnarray}
H_{\sigma\sigma\sigma\upsilon1} &=& \large[ H_x - (H_0 + H_{\sigma12} (\sigma_1 \sigma_2 + \sigma_2 \sigma_3) + H_{\sigma13} \sigma_1 \sigma_3 + H_{\sigma\upsilon1} \sigma_1 \upsilon_1 + H_{\sigma2\upsilon1} \sigma_2 \upsilon_1 \nonumber \\
&+&  H_{\sigma1\upsilon3} \sigma_3 \upsilon_1) \large ]/(\sigma_1 \sigma_2 \sigma_3 \upsilon_1)  \quad for \quad \upsilon_2 \rightarrow 0, \upsilon_3\rightarrow  0,\nonumber \\  
H_{\sigma\sigma\sigma\upsilon2} &=& \large[ H_x - (H_0 + H_{\sigma12} (\sigma_1 \sigma_2 + \sigma_2 \sigma_3) + H_{\sigma13} \sigma_1 \sigma_3 + 
        H_{\sigma\upsilon2} \sigma_2 \upsilon_2 + H_{\sigma1\upsilon2} \sigma_1 \upsilon_2  \nonumber \\
&+& H_{\sigma1\upsilon2} \sigma_3 \upsilon_2) \large]/(\sigma_1 \sigma_2 \sigma_3 \upsilon_2),  \quad for \quad \upsilon_1 \rightarrow 0, \upsilon_3\rightarrow  0,\nonumber \\
H_{\sigma1\upsilon\upsilon\upsilon} &=& \large[ H_x - (H_0 + H_{\sigma\upsilon1} \sigma_1 \upsilon_1 + H_{\sigma1\upsilon2} \sigma_1 \upsilon_2 + H_{\sigma1\upsilon3} \sigma_1 \upsilon_3 + H_{\upsilon12} \upsilon_1 \upsilon_2 +
    H_{\upsilon13} \upsilon_1 \upsilon_3 \nonumber \\
&+& H_{\upsilon12} \upsilon_2 \upsilon_3 \large] /(\sigma_1 \upsilon_1 \upsilon_2 \upsilon_3),  \quad for \quad \sigma_2 \rightarrow 0, \sigma_3\rightarrow  0, \nonumber \\
H_{\sigma2\upsilon\upsilon\upsilon} &=& \large[ H_x - (H_0 + H_{\sigma\upsilon2} \sigma_2 \upsilon_2 + H_{\sigma2\upsilon1} \sigma_2 \upsilon_1 +  H_{\sigma2\upsilon1} \sigma_2 \upsilon_3  
+ H_{\upsilon12} (\upsilon_1 \upsilon_2 + \upsilon_2 \upsilon_3)  \nonumber \\
&+& H_{\upsilon13} \upsilon_1 \upsilon_3 \large]/(\sigma_2 \upsilon_1 \upsilon_2 \upsilon_3),  \quad for \quad \sigma_1 \rightarrow 0, \sigma_3\rightarrow  0,\nonumber \\
H_{\sigma1\sigma2\upsilon1\upsilon2} &=& 
  \large[ H_x - (H_0 + H_{\sigma12} \sigma_1 \sigma_2 + H{_\sigma\upsilon1} \sigma_1 \upsilon_1 + H_{\sigma1\upsilon2} \sigma_1 \upsilon_2 + 
      H_{\sigma2\upsilon1} \sigma_2 \upsilon_1 + H_{\sigma\upsilon2} \sigma_2 \upsilon_2 \nonumber \\
&+& H_{\upsilon12} \upsilon_1 \upsilon_2 \large]/(\sigma_1 \sigma_2 \upsilon_1 \upsilon_2),  \quad for \quad \sigma_3 \rightarrow 0, \upsilon_3\rightarrow  0, \nonumber \\
H_{\sigma1\sigma2\upsilon2\upsilon3} &=& 
  \large[ H_x- (H_0 + H_{\sigma12} \sigma_1 \sigma_2 + H_{\sigma1\upsilon2} \sigma_1 \upsilon_2 + H_{\sigma1\upsilon3} \sigma_1 \upsilon_3 + 
      H_{\sigma\upsilon2} \sigma_2 \upsilon_2 + H_{\sigma2\upsilon1} \sigma_2 \upsilon_3 \nonumber \\
 &+& H_{\upsilon12} \upsilon_2 \upsilon_3 \large] /(\sigma_1 \sigma_2 \upsilon_2 \upsilon_3),  \quad for \quad \sigma_3 \rightarrow 0, \upsilon_1\rightarrow  0, \nonumber \\
H_{\sigma1\sigma2\upsilon1\upsilon3} &=& 
  \large[ H_x - (H_0 + H_{\sigma12} \sigma_1 \sigma_2 + H_{\sigma\upsilon1} \sigma_1 \upsilon_1 + H_{\sigma1\upsilon3} \sigma_1 \upsilon_3 + 
      H_{\sigma2\upsilon1} \sigma_2 \upsilon_1 + H_{\sigma2\upsilon1} \sigma_2 \upsilon_3 \nonumber \\
&+& H_{\upsilon13} \upsilon_1 \upsilon_3 \large] /(\sigma_1 \sigma_2 \upsilon_1 \upsilon_3),  \quad for \quad \sigma_3 \rightarrow 0, \upsilon_2\rightarrow  0, \nonumber \\
H_{\sigma1\sigma3\upsilon1\upsilon2} &=& 
  \large[ H_x - (H_0 + H_{\sigma13} \sigma_1 \sigma_3 + H_{\sigma\upsilon1} \sigma_1 \upsilon_1 + H_{\sigma1\upsilon2} \sigma_1 \upsilon_2 + 
      H_{\sigma1\upsilon3} \sigma_3 \upsilon_1 + H_{\sigma1\upsilon2} \sigma_3 \upsilon_2  \nonumber \\
&+& H_{\upsilon12} \upsilon_1 \upsilon_2\large] /(\sigma_1 \sigma_3 \upsilon_1 \upsilon_2 \large],   \quad for \quad \sigma_2 \rightarrow 0, \upsilon_3\rightarrow  0,  \nonumber \\ 
H_{\sigma1\sigma3\upsilon1\upsilon3} &=&  
\large[ H_x - (H_0 + H_{\sigma13} \sigma_1 \sigma_3 + H_{\sigma\upsilon1} \sigma_1 \upsilon_1 + H_{\sigma1\upsilon3} \sigma_1 \upsilon_3 + 
      H_{\sigma1\upsilon3} \sigma_3 \upsilon_1 + H_{\sigma\upsilon1} \sigma_3 \upsilon_3 \nonumber \\
&+& H_{\upsilon13} \upsilon_1 \upsilon_3 \large]/(\sigma_1 \sigma_3 \upsilon_1 \upsilon_3),  \quad for \quad \sigma_2 \rightarrow 0, \upsilon_2\rightarrow  0,  \nonumber \\
H_{\sigma\upsilon6} &=&  \large[ H_x- (H_0 + H_{\sigma12} (\sigma_1 \sigma_2 + \sigma_2 \sigma_3) + H_{\sigma13} \sigma_1 \sigma_3 +  H_{\upsilon12} (\upsilon_1 \upsilon_2 + \upsilon_2 \upsilon_3) \nonumber \\
&+& H_{\upsilon13} \upsilon_1 \upsilon_3 + H_{\sigma\upsilon1} (\sigma_1 \upsilon_1 + \sigma_3 \upsilon_3)   
+ H_{\sigma\upsilon2} \sigma_2 \upsilon_2 + H_{\sigma1\upsilon2} (\sigma_1 \upsilon_2 + \sigma_3 \upsilon_2) \nonumber \\
&+& H_{\sigma1\upsilon3} (\sigma_1 \upsilon_3 + \sigma_3 \upsilon_1) 
+  H_{\sigma2\upsilon1} (\sigma_2 \upsilon_1 + \sigma_2 \upsilon_3) + H_{\sigma\sigma\sigma\upsilon1} \sigma_1 \sigma_2 \sigma_3 (\upsilon_1 + \upsilon_3) \nonumber \\
&+&  H_{\sigma\sigma\sigma\upsilon2} \sigma_1 \sigma_2 \sigma_3 \upsilon_2 + H_{\sigma1\upsilon\upsilon\upsilon} (\sigma_1 + \sigma_3) \upsilon_1 \upsilon_2 \upsilon_3 
+ H_{\sigma2\upsilon\upsilon\upsilon} \sigma_2 \upsilon_1 \upsilon_2 \upsilon_3 \nonumber \\
 &+& H_{\sigma1\sigma2\upsilon1\upsilon2} (\sigma_1 \sigma_2 \upsilon_1 \upsilon_2 + \sigma_2 \sigma_3 \upsilon_2 \upsilon_3) 
+ H_{\sigma1\sigma2\upsilon2\upsilon3} (\sigma_1 \sigma_2 \upsilon_2 \upsilon_3 + \sigma_2 \sigma_3 \upsilon_1 \upsilon_2) \nonumber \\
&+& H_{\sigma1\sigma2\upsilon1\upsilon3} (\sigma_1 \sigma_2 + \sigma_2 \sigma_3) \upsilon_1 \upsilon_3 
+H_{\sigma1\sigma3\upsilon1\upsilon2} \sigma_1 \sigma_3 (\upsilon_1 \upsilon_2 + \upsilon_2 \upsilon_3) \nonumber \\
&+& H_{\sigma1\sigma3\upsilon1\upsilon3} \sigma_1 \sigma_3 \upsilon_1 \upsilon_3 \large] /(\sigma1 \sigma2 \sigma3 \upsilon1 \upsilon2 \upsilon3). \nonumber
\end{eqnarray}

Let us introduce the following notations
\begin{eqnarray}
Q_1 &=& H_x \quad for \quad {\sigma_1 \rightarrow 1, \sigma_2 \rightarrow 1, \sigma_3 \rightarrow 1, \upsilon_1 \rightarrow 1, \upsilon_2 \rightarrow 1, \upsilon_3 \rightarrow 1} \nonumber \\  
Q_2 &=& H_x \quad for \quad {\sigma_1 \rightarrow 1, \sigma_2 \rightarrow 1, \sigma_3 \rightarrow 1, \upsilon_1 \rightarrow 1, \upsilon_2 \rightarrow 1, \upsilon_3 \rightarrow -1} \nonumber \\  
Q_3 &=& H_x \quad for \quad {\sigma_1 \rightarrow 1, \sigma_2 \rightarrow 1, \sigma_3 \rightarrow 1, \upsilon_1 \rightarrow 1, \upsilon_2 \rightarrow -1, \upsilon_3 \rightarrow 1} \nonumber \\  
Q_4 &=& H_x \quad for \quad {\sigma_1 \rightarrow 1, \sigma_2 \rightarrow 1, \sigma_3 \rightarrow -1, \upsilon_1 \rightarrow 1, \upsilon_2 \rightarrow 1, \upsilon_3 \rightarrow 1} \nonumber \\  
Q_5 &=& H_x \quad for \quad {\sigma_1 \rightarrow 1, \sigma_2 \rightarrow -1, \sigma_3 \rightarrow 1, \upsilon_1 \rightarrow 1, \upsilon_2 \rightarrow 1, \upsilon_3 \rightarrow 1} \nonumber \\  
Q_6 &=& H_x \quad for \quad {\sigma_1 \rightarrow 1, \sigma_2 \rightarrow 1, \sigma_3 \rightarrow 1, \upsilon_1 \rightarrow 1, \upsilon_2 \rightarrow -1, \upsilon_3 \rightarrow -1} \nonumber \\  
Q_7 &=& H_x \quad for \quad {\sigma_1 \rightarrow 1, \sigma_2 \rightarrow 1, \sigma_3 \rightarrow 1, \upsilon_1 \rightarrow -1, \upsilon_2 \rightarrow 1, \upsilon_3 \rightarrow -1} \nonumber \\  
Q_8 &=& H_x \quad for \quad {\sigma_1 \rightarrow 1, \sigma_2 \rightarrow -1, \sigma_3 \rightarrow -1, \upsilon_1 \rightarrow 1, \upsilon_2 \rightarrow 1, \upsilon_3 \rightarrow 1} \nonumber \\  
Q_9 &=& H_x \quad for \quad {\sigma_1 \rightarrow -1, \sigma_2 \rightarrow 1, \sigma_3 \rightarrow -1, \upsilon_1 \rightarrow 1, \upsilon_2 \rightarrow 1, \upsilon_3 \rightarrow 1} \nonumber \\  
Q_{10} &=& H_x \quad for \quad {\sigma_1 \rightarrow 1, \sigma_2 \rightarrow 1, \sigma_3 \rightarrow -1, \upsilon_1 \rightarrow 1, \upsilon_2 \rightarrow 1, 
    \upsilon_3 \rightarrow -1} \nonumber \\  
Q_{11} &=& H_x \quad for \quad {\sigma_1 \rightarrow 1, \sigma_2 \rightarrow -1, \sigma_3 \rightarrow 1, \upsilon_1 \rightarrow 1, \upsilon_2 \rightarrow 1, 
    \upsilon_3 \rightarrow -1} \nonumber \\  
Q_{12} &=& H_x \quad for \quad {\sigma_1 \rightarrow -1, \sigma_2 \rightarrow 1, \sigma_3 \rightarrow 1, \upsilon_1 \rightarrow 1, \upsilon_2 \rightarrow 1, 
    \upsilon_3 \rightarrow -1} \nonumber \\  
Q_{13} &=& H_x \quad for \quad {\sigma_1 \rightarrow 1, \sigma_2 \rightarrow 1, \sigma_3 \rightarrow -1, \upsilon_1 \rightarrow 1, \upsilon_2 \rightarrow -1, 
    \upsilon_3 \rightarrow 1} \nonumber \\  
Q_{14} &=& H_x \quad for \quad {\sigma_1 \rightarrow 1, \sigma_2 \rightarrow -1, \sigma_3 \rightarrow 1, \upsilon_1 \rightarrow 1, \upsilon_2 \rightarrow -1, \upsilon_3 \rightarrow 1} \nonumber \\  
Q_{15} &=& H_x \quad for \quad {\sigma_1 \rightarrow -1, \sigma_2 \rightarrow 1, \sigma_3 \rightarrow 1, \upsilon_1 \rightarrow 1, \upsilon_2 \rightarrow -1, 
    \upsilon_3 \rightarrow -1} \nonumber \\  
Q_{16} &=& H_x \quad for \quad {\sigma_1 \rightarrow 1, \sigma_2 \rightarrow -1, \sigma_3 \rightarrow 1, \upsilon_1 \rightarrow 1, \upsilon_2 \rightarrow -1, 
    \upsilon_3 \rightarrow -1} \nonumber \\  
Q_{17} &=& H_x \quad for \quad {\sigma_1 \rightarrow 1, \sigma_2 \rightarrow 1, \sigma_3 \rightarrow -1, \upsilon_1 \rightarrow 1, \upsilon_2 \rightarrow -1, 
    \upsilon_3 \rightarrow -1} \nonumber \\  
Q_{18} &=& H_x \quad for \quad {\sigma_1 \rightarrow -1, \sigma_2 \rightarrow 1, \sigma_3 \rightarrow 1, \upsilon_1 \rightarrow -1, \upsilon_2 \rightarrow 1, 
    \upsilon_3 \rightarrow -1} \nonumber \\  
Q_{19} &=& H_x \quad for \quad {\sigma_1 \rightarrow 1, \sigma_2 \rightarrow -1, \sigma_3 \rightarrow 1, \upsilon_1 \rightarrow -1, \upsilon_2 \rightarrow 1, 
    \upsilon_3 \rightarrow -1} \nonumber \\  
Q_{20} &=& H_x \quad for \quad {\sigma_1 \rightarrow -1, \sigma_2 \rightarrow -1, \sigma_3 \rightarrow -1, \upsilon_1 \rightarrow 1, \upsilon_2 \rightarrow 1, 
    \upsilon_3 \rightarrow 1}. 
\end{eqnarray}
and 
\begin{equation}
q_i = \ln Q_i.
\end{equation}

All interaction parameters are expressed by the quantities $q_i$ however, we quote explicitly only those that were used in our calculations. So, the recursion relations for the renormalized interaction parameters (11) have the form
\begin{eqnarray}
Z_0 &=& \frac{1}{32} (q_1 + 2 q_{10} + 2 q_{11} + 2 q_{12} + 2 q_{13} + q_{14} + 2 q_{15} + 2 q_{16} + 
    2 q_{17} + 2 q_{18} + q_{19} + 2 q_2 \nonumber \\
&+& q_{20} + q_3 + 2 q_4 + q_5 + 2 q_6 + q_7 + 
    2 q_8 + q_9), \nonumber \\
J_1 &=& \frac{1}{32}(q_1 - 2 q_{11} - q_{14} - 2 q_{16} - q_{19} + 2 q_2 + q_{20}+q_3 - q_5 + 2 q_6 + q_7 - q_9), \nonumber \\
J_2 &=& 
\frac{1}{32} (q_1 - 2 q_{10} + 2 q_{11} - 2 q_{12} - 2 q_{13} + q_{14} - 2 q_{15} + 2 q_{16} -
    2 q_{17} - 2 q_{18} + q_{19} + 2 q_2 \nonumber \\
&+& q_{20} + q_3 - 2 q_4 + q_5 + 2 q_6 + q_7 -  2 q_8 + q_9), \nonumber \\
K_1 &=& 
 \frac{1}{32} (q_1 - 2 q_{13} - q_{14} - 2 q_{18} - q_{19} + q_{20} - q_3 + 2 q_4 + q_5 - q_7 + 2 q_8 + q_9), \nonumber \\
 K_2 &=& \frac{1}{32} (q_1 - 2 q_{10} - 2 q_{11} - 2 q_{12} + 2 q_{13} + q_{14} - 2 q_{15} - 2 q_{16} - 
    2 q_{17} + 2 q_{18} + q_{19} - 2 q_2 \nonumber \\
&+& q_{20} + q_3 + 2 q_4 + q_5 - 2 q_6 + q_7 +  2 q_8 + q_9), \nonumber \\
M_{01} &=& 
\frac{1}{32} (q_1 + 2 q_{10} - 2 q_{12} + q_{14} - 2 q_{15} + 2 q_{17} - q_{19} - q_{20} + q_3 + q_5 - q_7 - q_9), \nonumber \\
M_{02} &=& \frac{1}{32} (q_1 + 2 q_{10} - 2 q_{11} + 2 q_{12} - 2 q_{13} + q_{14} - 2 q_{15} + 2 q_{16} - 
    2 q_{17} + 2 q_{18} - q_{19} + 2 q_2 \nonumber \\
&-& q_{20} - q_3 + 2 q_4 - q_5 - 2 q_6 + q_7 -  2 q_8 + q_9), \nonumber \\
M_{1S }&=& \frac{1}{32} (q_1 + 2 q_{13} - q_{14} - 2 q_{18} + q_{19} - q_{20} + q_3 + 2 q_4 - q_5 - q_7 - 
    2 q_8 + q_9), \nonumber \\
M_{1V} &=& \frac{1}{32} (q_1 + 2 q_{11} - q_{14} - 2 q_{16} + q_{19} + 2 q_2 - q_{20} - q_3 + q_5 - 2 q_6 + 
    q_7 - q_9), \nonumber \\
M_2 &=& 
 \frac{1}{32} (q_1 - 2 q_{10} + 2 q_{12} + q_{14} + 2 q_{15} - 2 q_{17} - q_{19} - q_{20} + q_3 +  q_5 - q_7 - q_9).
\end{eqnarray}

%%%%%%%%%%%%%%%%%%%%%%%%%%%%%%%%%%%%%%%%%%k%%%%%%%%%%%%%%%%%%%%%%%%


\begin{thebibliography}{}
%%%%%%%%%%%%%%%%%%%%%%%%%%%%%%%%%%%%%%%%%%%%%%%%%%%%%%%%%%%%%%%%%%%
\bibitem{SW1} Sznajd-Weron K, Sznajd J 2005 Physica A {\bf 351} 593
\bibitem{Holyst}Ho\l yst J A, Kacperski K, Schweitzer F 2001 Annual Review of Computational Physics IX, World
Scientific, Singapore,  p 275 .
\bibitem{SW2} Sznajd-Weron K,  Sznajd J 2000 Int. J. Mod. Phys. C {\bf 11} 1157 
\bibitem{AT} Ashkin J, Teller E 1943  Phys. Rev. {\bf 64} 178 
\bibitem{Wen} Wen, J -J, Garlea V O, Koohpayeh S M, McQueen T M, Li H -F, Yan J -Q, Rodriguez-Rivera J A, Vaknin D, and Broholm C L 2015 Phys. Rev. B {\bf 91}, 054424
\bibitem{BS} Behera L and Schweitzer F 2003 Int. J. Mod. Phys. C {\bf 14} 1331
\bibitem{Fort} Castellano C, Fortunato S, Loreto V 2009  Reviews of Modern Physics {\bf 81} 591
\bibitem{TT} Toral R and Tessone C J 2007 Commun. Comput. Phys. {\bf 2} 177 
\bibitem{Stauffer1} Stauffer D, Sousa A O and Moss de Oliveira S 2000 Int. J. Mod. Phys. C {\bf 11} 1239
\bibitem{Stauffer2} Stauffer, D 2001 J. of Artificial Societies and Social Simulation, {\bf 5} no.1
\bibitem{Kasia1} Sznajd-Weron K 2002 Phys. Rev. E {\bf 66} 046131
\bibitem{Kasia2} Sznajd-Weron K 2004 Phys. Rev. E {\bf 70}  037104 
\bibitem{Galam} Galam Serge 2012, Sociophysics: A physicist's Modeling of Psycho-political Phenomena, Springer New York Dordrecht Heidelberg London

\end{thebibliography}
\end{document}